# Non-equilibrium Capacitance Loss in EDLC Based on Ion Flux

*Jia Li, Xinyu Zhang, Wei Wu\**

Laboratory of Printable Functional Materials and Printed Electronics, School of Physics and Technology, Wuhan University, Wuhan 430072, P. R. China

\* Corresponding author:

weiwu@whu.edu.cn (W. Wu)

**ABSTRACT.** Understanding the loss of non-equilibrium capacitance in the electric double layer capacitor (EDLC) during continuous charging compared to equilibrium capacitance is crucial for the design of high-rate energy storage devices. Due to three major challenges, including strong ion correlations induced by high concentrations at the interface, the difficulty in representing far from equilibrium, and numerical stability issues, current numerical simulations of ion distribution based on statistical mechanics struggle to achieve long-term accurate modeling under realistic parameters. Since non-equilibrium capacitance loss is a universal macroscopic phenomenon in EDLC, which exists independently of the microscopic model of the electrolyte, this study starts from macroscopic non-equilibrium phenomena, introduces a novel perspective centered on ion flux (IF) for the first time. By deriving the relationship among the charging current, IF, and displacement current, a novel capacitance model based on IF is first established. It is further demonstrated that the current relationship and voltage relationship of EDLC, when represented by conventional circuit models, exhibit inherent contradictions. Therefore, a new interpretative framework is proposed, wherein IF and displacement current are analogized to acceleration and deformation in an elastic medium, respectively. Inspired by this insight, this study reveals that the decrease in non-equilibrium capacitance with increasing charging current originates from the inherent resistance of IF to external charging. This tendency to restore equilibrium is the fundamental cause of non-equilibrium capacitance loss.

## I. INTRODUCTION

Continuous charging of the electric double layer capacitor (EDLC), such as galvanostatic charge/discharge (GCD) and cyclic voltammetry (CV) implies that thermodynamic equilibrium of the EDL cannot be established during this process. Both theoretical and experimental studies have shown that the differential capacitance of the EDLC under static conditions typically exhibits bell-shaped or camel-shaped behavior[1], whereas the CV curve measured under dynamic conditions usually appears quasi-rectangular[2]. Since non-equilibrium capacitance loss in EDLC occurs compared to equilibrium capacitance, and the loss increases with the magnitude of the charging current[2], revealing the dynamic mechanisms in EDLC charging and the principle of capacitance loss is crucial to providing valuable guidance for the design of high-rate energy storage devices.

EDLC charging arises from the microscopic motion of anions and cations under the external electric field, which leads to macroscopic evolution of the electrolyte, such as an increase (or decrease) in counterion (or co-ion) concentration[3]. The motion trajectories of a large number of ions in the electric field become complex due to frequent collisions[4]. Consequently, on a macroscopic time scale, the motion of individual ions in the electrolyte is subjected a combination of directional electric field and random collisions, resulting in drift-diffusion motion[5]. Due to the identical physical properties of cations (anions), at macroscopic time scales, the collective behavior of a large number of ions in the electrolyte manifests as the statistical average of the random motion of individual ions[6]. This implies that the macroscopic evolution of the electrolyte can be described without accounting for the informational redundancy introduced by the trajectories of individual ions.

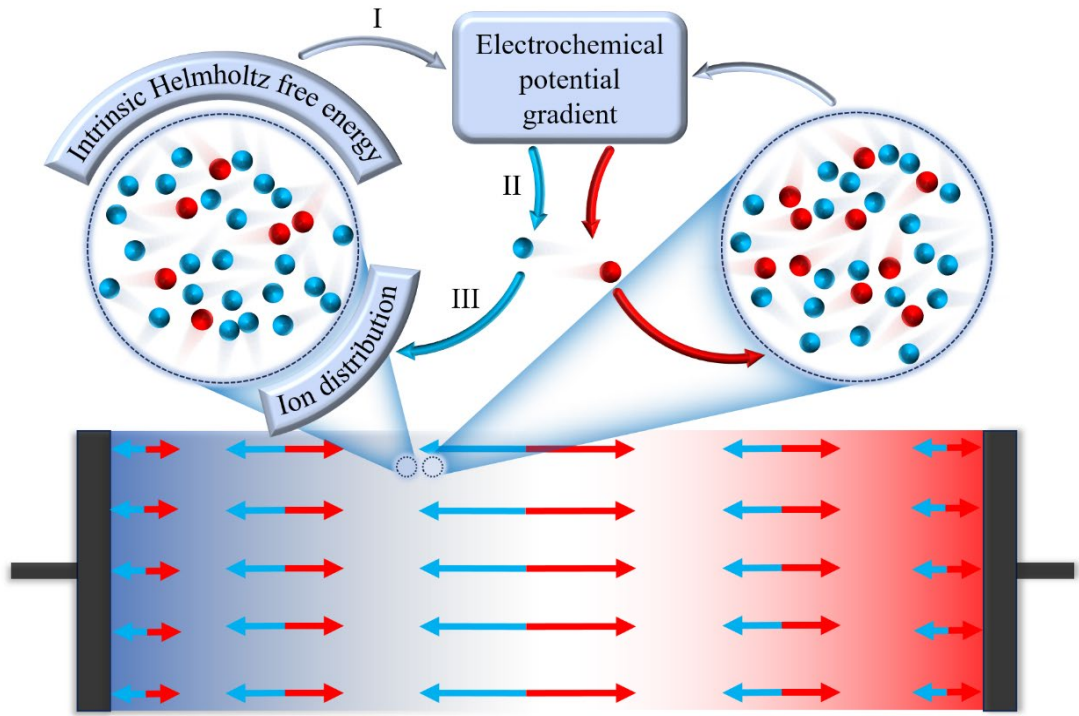

**FIG 1** The process of linking microscopic ion behavior to the capacitance of EDLC involves three major challenges, including local equilibrium statistical mechanics (I), phenomenological relation from electrochemical potential gradient to IF (II), solution of the Poisson–Nernst–Planck equation (III).

Starting from local equilibrium statistical mechanics, the electrochemical potential is derived from the intrinsic Helmholtz free energy of the electrolyte. By combining this with the phenomenological relations between thermodynamic fluxes and forces[7], the continuity equation describing ion number conservation, and the Poisson equation accounting for electrostatic screening, the widely used modified Poisson–Nernst–Planck (PNP) equation for studying transport processes is obtained. The ion and electric field distributions derived from the PNP model under IF and electric potential boundary conditions are used to calculate the non-equilibrium capacitance (Fig 1). This approach, however, is subject to three major challenges.

The first challenge is the strong correlation effects between ions, which is induced by the complex microscopic structure of the electrolyte and the high ion concentration at the interface, remain a fundamental challenge in statistical mechanics. Under strong ion-ion correlations, the ion-ion interactions within the macroscopic infinitesimal elements of the continuums can no longer be neglected, rendering the mean field theory inapplicable. As a result, more advanced methods of statistic mechanics have been developed to account for ion-ion correlations, surpassing the simpler and more convenient lattice gas models[8,9] For long-range interactions (e.g., Coulomb potential $1/r$), the inverse of Green's function exists, enabling the Hubbard-Stratonovich (HS) transformation within the field-theoretic (FT) approach to convert the partition function in ensemble statistics into a path integral. However, at high concentrations, ion-ion correlations induced by long-range interactions become significant and cannot be ignored, it is necessary to make beyond mean field corrections to the free energy functional of the electrolyte obtained from the saddle-point approximation[10], including mathematical treatments of the path integral through reference systems (e.g., Debye-Hückel interaction) using loop expansion[11], variational perturbation methods[12], and the GFB variational method[13]. For short-range interactions (e.g., Van der Waals interactions $1/r^6$), the inverse of Green's function required for the HS transformation does not exist[14]. Additionally, short-range interactions induce strong ion-ion correlations, which remain non-negligible even at low concentrations. As a result, classical density functional theory (cDFT) is the primary approach for handling such interactions. In this framework, the excess Helmholtz free energy functional, expressed in terms of the macroscopic two-body density function of the electrolyte, is approximated using reference systems (e.g., hard-sphere fluids) via density expansion. The second-order expansion is typically treated using the square gradient approximation (SGA), which requires solving the Ornstein-Zernike (OZ) integral equation with closure relations such as the Hypernetted Chain (HNC) approximation[15,16] and the mean spherical approximation (MSA)[17] to obtain the direct correlation functions (DCF). However, the computational complexity of solving the OZ integral equation and the difficulty in achieving accurate closure approximations[18] under strong ion-ion correlation effects pose significant challenges for statistical mechanics in accurately describing the ion-ion correlations in high-concentration, non-ideal electrolytes at the EDL.

The second challenge lies in the fact that the PNP equation is not suitable for studying interfacial transport phenomena with far from equilibrium (such as high current charging), as the local equilibrium approximation only applies to near equilibrium regime[19] and phenomenological relationships are strictly valid only for bulk phase transport processes[6,20]. The third challenge lies in solving the PNP equations, even when simplified to the linear Debye-Falkenhagen equation, it requires complex handling of the Laplace inverse transformation to obtain a satisfactory primitive function[21]. However, numerical solutions face multiple challenges. The nonlinear coupling between the electric field and concentration field[22] increases both the computational cost and complexity. Moreover, excessive concentration and potential gradient variations at the interface introduce rigid problems that decrease solution stability. Therefore, under the premise of ensuring the authenticity of physical parameters[9,23], achieving stable computation for EDLC during long-time charging poses significant challenges.

This suggests a reconsideration of the non-equilibrium phenomena in EDLC charging, as the direct solution object of the PNP model is the ion distribution, but the key to the non-equilibrium phenomenon is not the ion distribution, but the thermodynamic fluxes or thermodynamic forces. Moreover, non-equilibrium capacitance loss in EDLC is a universal macroscopic phenomenon that exists independently of the microscopic model of the electrolyte. Therefore, this study introduces for the first time a novel perspective centered on ion flux (IF). By deriving the relationship among the charging current, IF, and displacement current, a novel capacitance model based on IF is first established. It is further demonstrated that the current relationships and voltage relationships of EDLCs cannot be represented simultaneously within a unified circuit model. To address this, a new interpretation is proposed by analyzing the properties of IF, wherein the EDLC charging process is analogized to a mechanical process. Specifically, since the interfacial ion accumulation induced by IF hinders its own increase, the IF and the displacement current representing ion accumulation, are analogous to acceleration and deformation in an elastic medium, respectively. Inspired by this insight, analysis via the Poisson equation obtains that the increment of the total IF in the electrolyte gradually diminishes as the current increases. This founding reveals that the decrease in non-equilibrium capacitance with increasing charging current originates from the inherent resistance of IF to external charging. This tendency to restore equilibrium is the fundamental cause of non-equilibrium capacitance loss in EDLC.

## II. MODEL

The subject of this study is the electrode-electrolyte interface. To solely investigate the transport process of the electrolyte, it is necessary to assume that the electrode is a hard wall that completely blocks ions, while also considering the electrolyte as a medium that completely blocks electrons. Electrolyte systems are usually considered to be isotropic medium, ignoring the inductive electric field generated by IF during charging and only considering the electrostatic field. Therefore, ions only transport in the direction of the electrostatic field, while there is only thermal motion of ions in the vertical direction, with a macroscopic velocity of zero. Therefore, to investigate the one-dimensional ion transport process, the left electrode is defined as the cathode, and the right electrode as the anode. The direction of the charging current $I$ is positive, and a coordinate system is established with the midpoint between the two electrodes as the origin (Fig. 2).

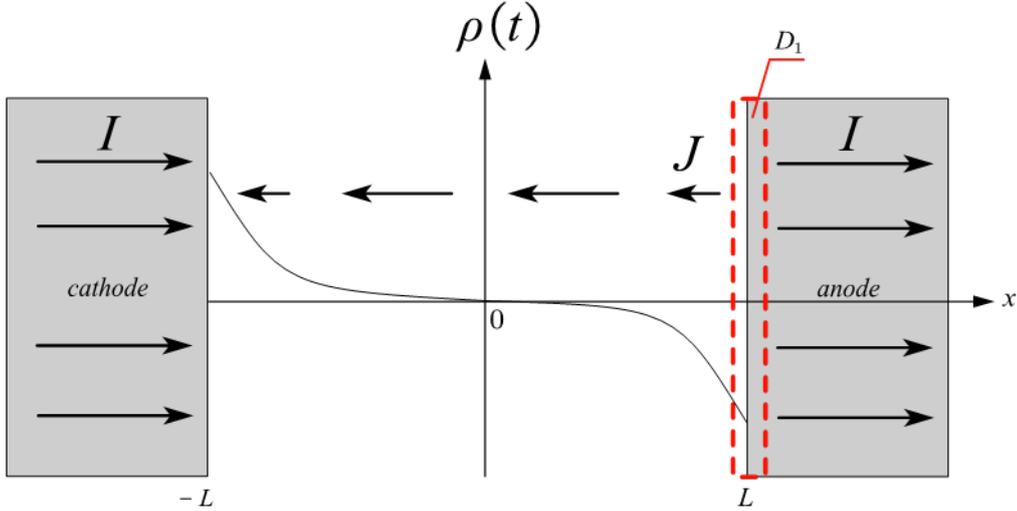

FIG 2 Schematic of the net ion spatial distribution

To simplify the discussion, the electrolyte is set as a symmetric type of $z-z$ ionic system, and the dielectric constant $\varepsilon_s$ of the electrolyte is uniform everywhere. Define the difference between anions and cations distribution as net ion distribution:

$$\rho \triangleq c_+ - c_- \tag{1}$$

Similarly, the difference between the cation and anion flux is defined as the net ion flux $J$, referred to as ion flux (IF):

$$J \triangleq J_+ - J_- \tag{2}$$

Due to the adsorption of counterions and

repulsion of co-ions by the charging electrode, the IF is opposite in direction to the charging current $I$, thus remaining negative in this model setup.

## III. RELATIONSHIP BETWEEN IF AND NON-EQUILIBRIUM CAPACITANCE IN EDLC

Based on the conservation of ion number and the combination with Gauss's law for electrostatic fields, this study concludes that the joint physical quantity of IF and displacement current has a non-divergent property, which means that in the electrolyte, this joint physical quantity is equal everywhere. Since the IF has a Neumann boundary condition that is always zero, the boundary condition for the displacement current can be derived using the charging current. Therefore, by utilizing the charging current, the relationship between the IF and the displacement current can be obtained. Since the displacement current symbolizes changes in device voltage and the current symbolizes changes in electronic charge capacity at the electrode, by using the definition of device capacitance, the relationship between IF and non-equilibrium capacitance in EDLC is obtained. The specific steps of the above process are as follows. (To verify the correctness of the results, a more common dipole moment is used as the starting point in Appendix A).

First, Gauss's law is used to describe the relationship between the electrostatic field $E$ and net ion distribution $\rho$:

$$\frac{\partial}{\partial x}E = \frac{ze}{\varepsilon_s}\rho \qquad (3)$$

where $e$ represents the elementary electric charge. Then, according to the conservation of cation number and anion number, the conservation of net ion number is obtained:

$$\frac{\partial \rho}{\partial t} = -\frac{\partial J}{\partial x} \qquad (4)$$

With Eqs. (3) and (4), we obtain the following equation:

$$\frac{\partial}{\partial x}\left(zeJ + \varepsilon_s \frac{\partial E}{\partial t}\right) = 0 \qquad (5)$$

The equation shows that the sum of current density by IF and displacement currents $\varepsilon_s \partial E/\partial t$ is equal everywhere. Therefore, so long as the values of IF and displacement currents at the boundary are known, the specific value of this sum can be obtained. Due to the blocking effect of hard-wall electrodes on ions, the IF at the boundary is always zero[24,25]. The displacement current at the boundary is caused by electrode charging, for which we take an arbitrarily small closed area $D_1$ on the surface of the anode (Fig 2), and the relationship between its electric flux and the electronic charge capacity at the anode surface $Q_{anode}$ is given by the integral form of Gauss's law[26]:

$$\oint_{D_1} \varepsilon \mathbf{E} \cdot d\mathbf{S} = Q_{anode} \qquad (6)$$

Since the inner electric field of electrode is approximately zero, it can be deduced from the above equation that the electric field at the anode boundary of the ion system:

$$E\big|_{x=L} = -\frac{Q_{anode}}{\varepsilon_s S} \qquad (7)$$

Where $x=L$ is the anode boundary of the ionic system, $S$ represents the contact area between the electrode and electrolyte. Therefore, the summation of IF and displacement current at the anode boundary can be obtained:

$$\left(zeJ + \varepsilon_s \frac{\partial E}{\partial t}\right)\bigg|_{x=L} = -\frac{I}{S} \qquad (8)$$

Where $I$ represents the charging current. With Eq. (5) and boundary conditions Eq.(8), the relationship between charging current, IF, and displacement current is obtained:

$$-\varepsilon_s \frac{\partial E}{\partial t} = \frac{I}{S} + zeJ \qquad (9)$$

This formula indicates that due to ion accumulation, the electronic current on the electrode and the ionic current in the electrolyte do not satisfy a series connection. The difference between the two is equal to the displacement current induced by the ion accumulation, which in turn hinders the increase in IF.

Traditionally, the difference between the anode's potential and the cathode's potential is defined as the device voltage, which is precisely the potential difference across the electrolyte due to the continuity of potential at the boundaries[22]. According to the relationship of electrostatic field $E$ and potential $\varphi$: $E = -\nabla\varphi$, the device voltage is represented as:

$$U = \varphi(L) - \varphi(-L) = -\int_{-L}^{L} E(x)dx \qquad (10)$$

Therefore, by taking the integral of both sides of the Eq. (9), the left side yields the device voltage.

$$\frac{dU}{dt} = \frac{I}{\varepsilon_s S} 2L + \frac{ze}{\varepsilon_s} \int_{-L}^{L} J dx \qquad (11)$$

The capacitance of the device is defined as the rate of change the charge capacity on the anode $Q_{anode}$ with respect to the change in the device voltage $U$.

$$C = \frac{dQ_{anode}}{dU} = \frac{I}{\frac{dU}{dt}} \qquad (12)$$

Therefore, substituting the Eq. (11) into the definition of capacitance, the relationship between the IF and the non-equilibrium capacitance in EDLC is obtained:

$$\frac{1}{C} = \frac{1}{C_D} + \frac{ze \int_{-L}^{L} J dx}{\varepsilon_s \quad I} \qquad (13)$$

Where $C_D$ represents the capacitance contributed by the dielectric properties of the electrolyte: $C_D = \varepsilon_s S / 2L$, it is exactly the capacitance of a parallel-plate capacitor, This indicates that the capacitance of the EDLC is jointly composed of the dielectric capacitance and the influence introduced by the electrolyte. Importantly, the effect of the electrolyte is not manifested as an independent capacitance element, but rather through the generation of IF opposite to the charging current $I$, which leads to an overall the capacitance in EDLC greater than the capacitance in dielectric. Further explanation in Appendix A shows that the electrolyte functions by reducing the EDLC voltage via electrostatic screening, thereby increasing the capacitance. Additionally, Appendix B proves the existence of the inequality:

$$C_D \leq C < +\infty \qquad (14)$$

## IV. PROPERTIES AND APPLICATIONS OF IF

It is widely recognized that EDLC are typically modeled as a parallel circuit composed of dielectric and electrolyte. However, this study reveals inherent contradictions in this conventional understanding and proposes an alternative perspective for interpreting the EDLC charging process.

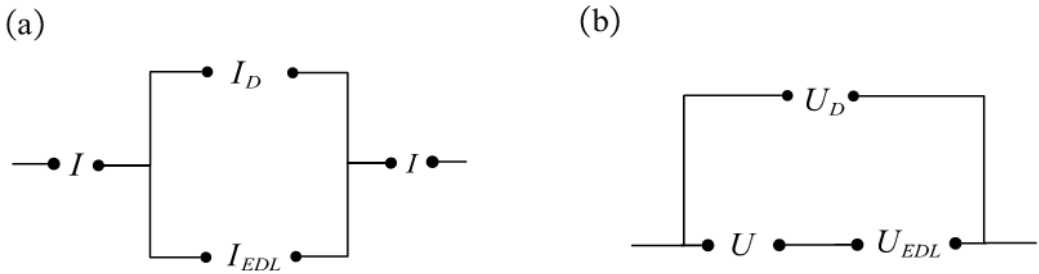

FIG 3 (a) Parallel connection of the displacement current in the dielectric and the ion current in the electrolyte resulting in the reverse electrode charging current; (b) Series connection of the EDLC voltage and the voltage reduction induced by the electrolyte yielding the voltage across the dielectric.

Since displacement current exists within the dielectric, $I_D$ is used to denote the displacement current in the opposite direction, while $I_{lyte}$ represents the ionic current in the electrolyte, also in the opposite direction (the negative sign arises from the convention that electronic current carries negative charge):

$$I_D \triangleq -\varepsilon_s S \frac{\partial E}{\partial t} \quad (15)$$

$$I_{lyte} \triangleq -zeSJ \quad (16)$$

The Eq. (9) can be simplified as: $I = I_D + I_{lyte}$, When this relation is represented in an equivalent circuit form, it corresponds to the circuit shown in Fig 3(a). From the perspective of current, the electrode current results from the parallel combination of the electrolyte and the dielectric. Moreover, according to the relationship among the EDLC voltage, dielectric layer voltage, and electrolyte voltage derived in Appendix A:

$$U = \frac{Q_{anode}}{\varepsilon_s S} 2L + \frac{ze}{\varepsilon_s} \int_{-L}^{L} x\rho dx \quad (17)$$

In the right-hand side of the above equation, the first term represents the voltage across the dielectric layer, while the second term corresponds to the voltage reduction induced by electrostatic screening from the electrolyte (i.e., $\int_{-L}^{L} x\rho dx$ is negative). These terms are hence denoted as:

$$U_D \triangleq \frac{Q_{anode}}{\varepsilon_s S} 2L \quad (18)$$

$$U_{lyte} \triangleq -\frac{ze}{\varepsilon_s} \int_{-L}^{L} x\rho dx \quad (19)$$

The Eq. (17) can be simplified as: $U_D = U + U_{lyte}$, When represented in an equivalent circuit form, it corresponds to the circuit shown in Fig 3(b). From the perspective of voltage, the EDLC voltage and the reverse voltage of the electrolyte are connected in series to yield the voltage across the dielectric. A comparison of Fig 3(a) and Fig 3(b) reveals a fundamentally contradictory series-parallel relationship, indicating that the electrical behavior of EDLC cannot be accurately captured by conventional circuit elements. This is because the electrolyte does not intrinsically behave as a resistor, capacitor, or their combination. As shown in the capacitance expression (Eq. (13)) and voltage expression (Eq. (17)), the electrolyte functions by reducing the EDLC voltage via dipole formation from ion separation, thereby enhancing capacitance, rather than acting as a capacitor-resistor-capacitor (CRC) unit in parallel with the dielectric. Furthermore, due to the pronounced diffusion of ions within the EDL, the electrolyte cannot be approximated by a parallel-plate capacitor composed of tightly packed ions. The inherent differences in the thermal motion of electrons and ions preclude a direct analogy between EDLC charging and conventional electronic circuits.

Therefore, the behavior of IF cannot be fully understood within the framework of conventional circuit theory. As IF inherently falls within the domain of non-equilibrium statistical mechanics, Appendix C discusses how ion drift under an electric field leads to interfacial ion accumulation, which in turn induces repulsion between like-charged ions (i.e., electrostatic screening) and induces counteracting diffusion flux, both effects serve to limit further increases in IF. A similar resistance to externally induced changes is observed in elastic media within mechanical systems. Given the well-established analogy between "electric potential differences" in circuits and "forces" in mechanical systems[27], this study further identifies an extended correspondence between statistical systems and mechanical systems, as summarized in the TABLE I.

Re-examining Eq. (9) reveals that the charging current, IF, and displacement current (representing ion accumulation) can be respectively analogized to the externally applied acceleration, the acceleration of the elastic medium, and the deformation of the elastic medium Under galvanostatic

**TABLE I** Comparison of the Mechanical Properties of Elastic Medium and the Statistical Properties of electrolyte.

|  | Mechanical Properties of Elastic Medium | Statistical Properties of electrolyte |
|---|---|---|
| Equilibrium | $a = 0$ | $J_\pm = 0$ |
| Non-Equilibrium | $a \propto -\nabla \sigma$ | $J_\pm \propto -\nabla \tilde{\mu}_\pm$ |
| Tendency to restore equilibrium | Deformation | Ion accumulation |

Note: $a$ represents acceleration field of the elastic medium; $\sigma$ represents the stress field of the elastic medium; $\tilde{\mu}_\pm$ represent the electrochemical potentials of cations and anions, respectively.

charging conditions, the IF in the bulk phase $(x = 0)$ reaches a steady value after the relaxation time $\varepsilon_s/\sigma$, expressed as follows (the detailed derivation is provided in Appendix B):

$$J\big|_{x=0} = -\frac{I}{zeS} \quad (20)$$

The above equation indicates that the charging current directly governs the IF in the bulk phase. From the perspective of IF, the EDLC charging process can be described as follows: the charging current applied at the electrode induces an IF in the bulk, resulting in non-equilibrium ion exchange within the electrolyte. Since the IF at the boundary is always zero, the non-uniform IF leads to ion accumulation. As previously discussed, the IF and ion accumulation are analogous to the acceleration and deformation of an elastic medium, respectively. Therefore, the galvanostatic charging process of the electrolyte can be analogized to the application of constant acceleration to an elastic medium (Fig 4), Similarly, the potentiostatic charging process can be analogized to the deformation of an elastic medium under a constant external force.

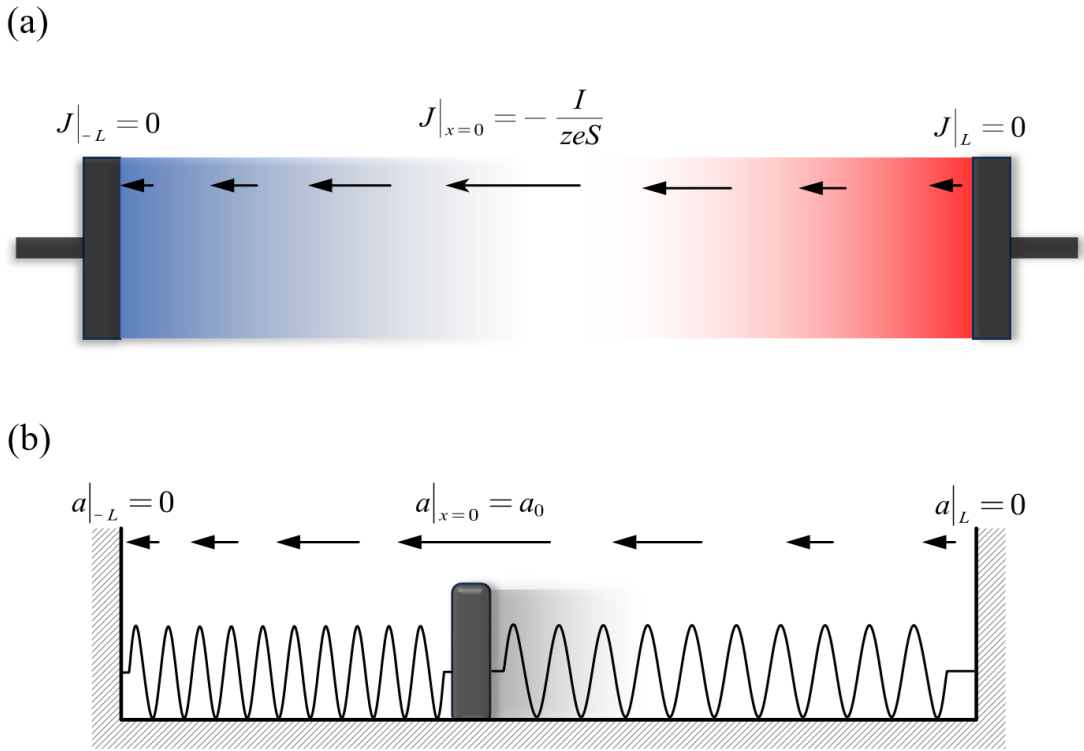

**FIG 4** (a) Galvanostatic charging applied to the electrolyte, (b) Constant acceleration applied to the elastic medium inducing deformation.

Since elastic medium resist external influences through deformation, thereby hindering the increase in its acceleration, it can be intuitively understood that the total acceleration of the medium will not increase linearly with the increase of the constant acceleration $a_0$ input by external. Similarly, the electrolyte will resist external influences through ion accumulation, thereby hindering the increase of IF in the electrolyte. Therefore, it can also be intuitively understood that the total IF in the electrolyte will not increase linearly with the increase of galvanostatic $I$. In addition, this process can also be qualitatively described using the follow process:

Galvanostatic charging maintains the electrolyte system in a continuous non-equilibrium state. Upon the removal of the current, the IF in the electrolyte gradually decays to zero. However, during this process, the value of net ion concentration $\rho$ at the interface continues to increase. According to the Poisson equation of the electrostatic field:

$$\nabla \cdot \mathbf{J} \propto -\frac{\partial^2}{\partial x^2}\varphi = \frac{ze}{\varepsilon_s}\rho \qquad (21)$$

The left side of equation represents the IF divergence caused by electrostatic interaction, so it can be explained that the non-equilibrium process will have a smaller value of IF divergence than the equilibrium process. Furthermore, the higher the degree of non-equilibrium, the smaller the value of the IF divergence. Therefore, a higher galvanostatic results in a sharper spatial distribution of the IF (Fig 5).

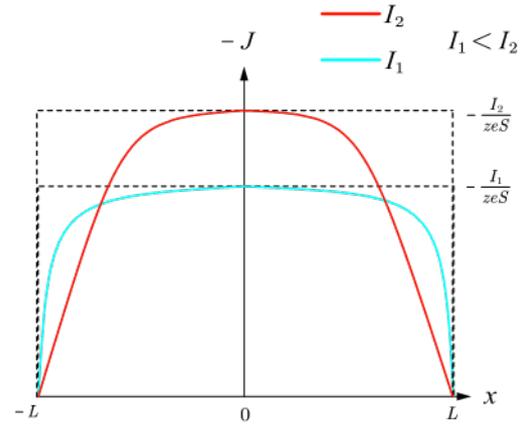

**FIG 5** Schematic diagram of the variation of IF with galvanostatic

The area in Fig 5 represents the $\int_{-L}^{L} Jdx$ in Eq. (13), Although $J|_{x=0}$ increases linearly with the galvanostatic, due to the shape becomes sharper, a qualitative relationship between the area and the galvanostatic can be established:

$$\int_{-L}^{L} Jdx \propto -I^a \ ; \ a<1 \qquad (22)$$

Substituting the above formula back into Eq. (13), a qualitative relationship between the non-equilibrium capacitance in EDLC and the galvanostatic is obtained:

$$\frac{1}{C} \propto \frac{1}{C_D} - \frac{ze}{\varepsilon_s}\frac{1}{I^{1-a}} \ ; \ a<1 \qquad (23)$$

Since it has been proven in Eq. (14) that the first term of the above formula is always greater than the second term, and given that $ze/\varepsilon_s > 0$, the larger galvanostatic $I$ applied, the smaller the non-equilibrium capacitance $C$ of the device. Therefore, since IF inherently resist external charging by inducing ion accumulation at the interface, which in turn hinders its own increase, this mechanism leads to the phenomenon where non-equilibrium capacitance in EDLC decreases with increasing charging current. This founding reveals that this tendency to restore equilibrium is the fundamental cause of non-equilibrium capacitance loss in EDLC.

## V. CONCLUSION

This study addresses the issue of capacitance loss induced by high current charging and, for the first time, introduces a novel perspective centered on IF, providing a fundamental physical explanation and understanding of the phenomenon. By leveraging the divergence-free characteristic exhibited by the combination of IF and displacement current, along with the boundary conditions of displacement current induced by charging, the relationship among charging current, IF, and displacement current is successfully derived, based on which a novel capacitance model expressed in terms of IF is established. It is further demonstrated that this model inherently excludes the occurrence of negative capacitance, and reveals that the current relations and voltage relations of EDLC correspond to different series-parallel configurations, this contradiction arises from the fact that the IF cannot be regarded as a conventional circuit element. Through analysis, this study reveals that both the electrolyte and the elastic medium exhibit resistance to externally induced changes. Accordingly, the IF and the displacement current representing ion accumulation, can be analogized to the acceleration and deformation of an elastic medium, respectively. Additionally, since the bulk IF quickly aligns with the galvanostatic, galvanostatic charging can be compared to the constant-acceleration stretching process of an elastic medium. Inspired by this insight, a qualitative analysis based on Poisson's equation leads to the conclusion that "although bulk IF increases with increasing galvanostatic, the increment of total IF in the electrolyte gradually weakens", This founding reveals that the decrease in non-equilibrium capacitance with increasing charging current originates from the inherent resistance of IF to external charging. This tendency to restore equilibrium is the fundamental cause of non-equilibrium capacitance loss in EDLC.

## CONFLICT OF INTEREST STATEMENTS

The authors declare that they have no known competing financial interests or personal relationships that could have appeared to influence the work reported in this paper.

## ACKNOWLEDGMENTS

This work was supported by the National Natural Science Foundation of China (NSFC, No. 52373252), National High-Level Talents Special Support Program.

## APPENDIX A: DERIVATION OF THE RELATIONSHIP BETWEEN IF AND NON-EQUILIBRIUM CAPACITANCE IN EDLC BASED ON LEIBNIZ INTEGRAL RULES

By applying Gauss's law (Eq. (A1)), which describes the relationship between the electrostatic field $E$ and net ion distribution $\rho$, along with the electric field at the electrolyte-electrode interface (Eq. (A2)).

$$\frac{\partial}{\partial x}E = \frac{ze}{\varepsilon_s}\rho \qquad (A1)$$

$$E\big|_{x=L} = -\frac{Q_{anode}}{\varepsilon_s S} \qquad (A2)$$

The electric field distribution (Eq.(A3)) can be obtained through Leibniz integration.

$$E(x) = -\frac{Q_{anode}}{\varepsilon_s S} - \frac{ze}{\varepsilon_s}\int_x^L \rho(x')dx' \qquad (A3)$$

Where $\rho(x')$ represents the net ion distribution at position $x'$. According to the relationship between the electrostatic field and electrostatic potential: $E = -\nabla\varphi$, thus, by performing Leibniz integration on the above equation once again, the potential difference across the electrolyte can be obtained. Conventionally, the potential difference between the anode and cathode is defined as the device voltage. Therefore, the device voltage $U$ can be expressed as:

$$\begin{aligned}
U &= -\int_{-L}^{L} E(x)dx \\
&= \int_{-L}^{L} \left( \frac{Q_{anode}}{\varepsilon_s S} + \frac{ze}{\varepsilon_s} \int_{x}^{L} \rho(x')dx' \right) dx \\
&= \frac{Q_{anode}}{\varepsilon_s S} 2L + \frac{ze}{\varepsilon_s} \left( \int_{x}^{L} (x-x')\rho(x')dx' \right)\bigg|_{x=-L}^{x=L} \quad (A4) \\
&= \frac{Q_{anode}}{\varepsilon_s S} 2L - \frac{ze}{\varepsilon_s} \int_{-L}^{L} (-L-x')\rho(x')dx' \\
&= \frac{Q_{anode}}{\varepsilon_s S} 2L + \frac{ze}{\varepsilon_s} \int_{-L}^{L} x\rho dx
\end{aligned}$$

The final step utilizes the assumption that the total net ion count in the electrolyte remains zero. The first term in the device voltage represents the voltage division across the dielectric, while the second term, where $ze\int_{-L}^{L} x\rho dx$ corresponds to the dipole moment generated by the separation of cations and anions, represents the voltage division of the EDL. Since these two terms have opposite signs, the voltage in EDLC is lower than the voltage in the dielectric. The physical origin of this conclusion lies in the fact that the charging process induces the separation of cations and anions within the ionic system, generating a dipole moment that electrostatic screening of the external electric field. This screening effect weakens the electric field within the ionic system, resulting in a device voltage lower than that of a corresponding parallel-plate capacitor. Then, by applying the definition of the device's non-equilibrium capacitance:

$$C = \frac{dQ_{anode}}{dU} = \frac{I}{\frac{dU}{dt}} \quad (A5)$$

Therefore. Substituting the expression for the device voltage (Eq. (A4)) into the above equation yields the relationship between the dipole moment and the device's non-equilibrium capacitance.

$$\frac{1}{C} = \frac{1}{C_D} + \frac{ze}{\varepsilon_s} \frac{\frac{d}{dt}\int_{-L}^{L} x\rho dx}{I} \quad (A6)$$

Where $C_D$ represents the capacitance induced by the dielectric constant of the electrolyte: $C_D = \varepsilon_s S/2L$, Therefore, electrostatic screening causes the voltage in EDLC to be lower than the voltage in the dielectric, resulting in the EDLC capacitance $C$ always being greater than the dielectric capacitance $C_D$. Since $\int_{-L}^{L} x\rho dx$ originates from the non-uniform transport of anions and cations, particularly the obstruction of IF at the electrode boundary, the dipole moment must inherently be related to the IF. First, the time derivative of the dipole moment is expanded according to Leibniz integral rule.

$$\frac{d}{dt}\int_{-L}^{L} x\rho dx = \int_{-L}^{L} \frac{\partial x}{\partial t}\rho dx + \int_{-L}^{L} x\frac{\partial \rho}{\partial t}dx \quad (A7)$$

Since the coordinate variable $x$ does not explicitly depend on time $t$, the first term on the right-hand side of the above equation is equal to zero. The second term in the above equation can be related to the IF through the conservation equation of net ion number:

$$\frac{\partial \rho}{\partial t} = -\frac{\partial J}{\partial x} \quad (A8)$$

Substituting this into the second term on the right-hand side of Eq. (A7) and applying integration by parts yields:

$$-\int_{-L}^{L} x\frac{\partial J}{\partial x}dx = -(Jx)\big|_{-L}^{L} + \int_{-L}^{L} Jdx \quad (A9)$$

Since the electrodes are considered to completely obstruct ionic motion, the IF at the boundary is always zero. Consequently, the relationship between the dipole moment and IF can be obtained.

$$\frac{d}{dt}\int_{-L}^{L} x\rho dx = \int_{-L}^{L} Jdx \quad (A10)$$

This equation indicates that the rate of change of the dipole moment is equal to the total IF in the electrolyte. Substituting this expression back into Eq. (A6) yields the relationship between IF and the non-equilibrium capacitance in EDLC.

$$\frac{1}{C} = \frac{1}{C_D} + \frac{ze}{\varepsilon_s} \frac{\int_{-L}^{L} Jdx}{I} \quad (A11)$$

## APPENDIX B: BEHAVIOR OF BULK IF UNDER GALVANOSTATIC CHARGING

Unlike the equilibrium state eventually reached under potentiostatic charging, in the case of galvanostatic charging, the continuous increase in charge capacity of the electrode, the ionic system must continuously respond to the ongoing external input, thus it cannot reach equilibrium. Instead, what state it ultimately achieve?

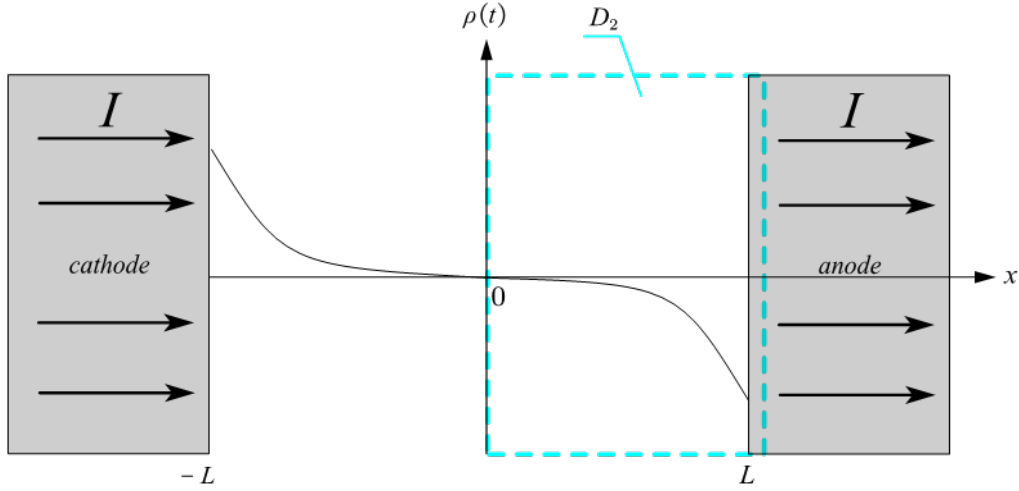

**FIG 6** The enclosed region spanning from the inner surface of the anode to the bulk phase of the electrolyte.

The transport process at the EDL consists of the electromigration and opposite-direction diffusion flux, involving the coupling of the electric field and concentration field, making it analytically intractable. In contrast, the transport process in bulk-phase ($x = 0$) of electrolyte (referred to as bulk) is governed solely by electromigration and can be simply described by the differential form of the phenomenological Ohm's law:

$$J|_{x=0} = \frac{\sigma}{ze} E|_{x=0} \quad (B1)$$

Where $\sigma$ represents bulk conductivity. Similarly, the bulk electric field $E|_{x=0}$ can be related to the charging conditions using the integral form of Gauss's law. As shown in Fig 6, the electric flux within the enclosed region from the anode to the bulk is given by:

$$\oint_{D_2} \varepsilon \mathbf{E} \cdot d\mathbf{S} = Q_{anode} + Q_{lyte} \quad (B2)$$

Where $Q_{lyte}$ represents the net ionic charge capacity from the anode to the bulk:

$$Q_{lyte} = zeS \int_0^L \rho dx \quad (B3)$$

For galvanostatic $I$ charging, the electronic charge capacity on the anode satisfies:

$$Q_{anode} = It \quad (B4)$$

Similarly, since the electric field inside the electrode is zero, the relationship between the bulk electric field and the applied galvanostatic can be established.

$$E|_{x=0} = -\frac{It + zeS \int_0^L \rho dx}{\varepsilon_s S} \quad (B5)$$

Substituting this into the bulk Ohm's law (Eq. (B1)) yields the bulk IF:

$$J|_{x=0} = -\frac{\sigma}{ze\varepsilon_s S}\left(It + zeS \int_0^L \rho dx\right) \quad (B6)$$

Taking the time derivative on both sides of the above equation yields:

$$\frac{d}{dt} J|_{x=0} = -\frac{\sigma I}{ze\varepsilon_s S} - \frac{\sigma}{\varepsilon_s} \frac{d}{dt} \int_0^L \rho dx$$
$$= -\frac{\sigma I}{ze\varepsilon_s S} - \frac{\sigma}{\varepsilon_s} \int_0^L \frac{\partial \rho}{\partial t} dx \quad (B7)$$

Therefore, substituting the conservation equation of net ion number (Eq. (A8)) into the above expression and considering that the IF at the boundary is always zero, the time-dependent equation for bulk IF can be obtained.

$$\frac{d}{dt}J\big|_{x=0} = -\frac{\sigma I}{ze\varepsilon_s S} - \frac{\sigma}{\varepsilon_s}J\big|_{x=0} \quad (B8)$$

Based on the natural initial condition: $J\big|_{x=0,t=0} = 0$, solving Eq. (B8) yields the time-dependent function for bulk IF.

$$J\big|_{x=0} = \frac{I}{zeS}e^{-\frac{\sigma}{\varepsilon_s}t} - \frac{I}{zeS} \quad (B9)$$

From this function, it can be observed that under galvanostatic charging, the magnitude of bulk IF increases over time until it reaches relaxation time $\varepsilon_s/\sigma$, after which it stabilizes and maintains a proportional relationship with the charging current (Fig 7(a)). Therefore, galvanostatic charging results in a unique state where the electrode current and the bulk IF attain equal magnitude but opposite direction.

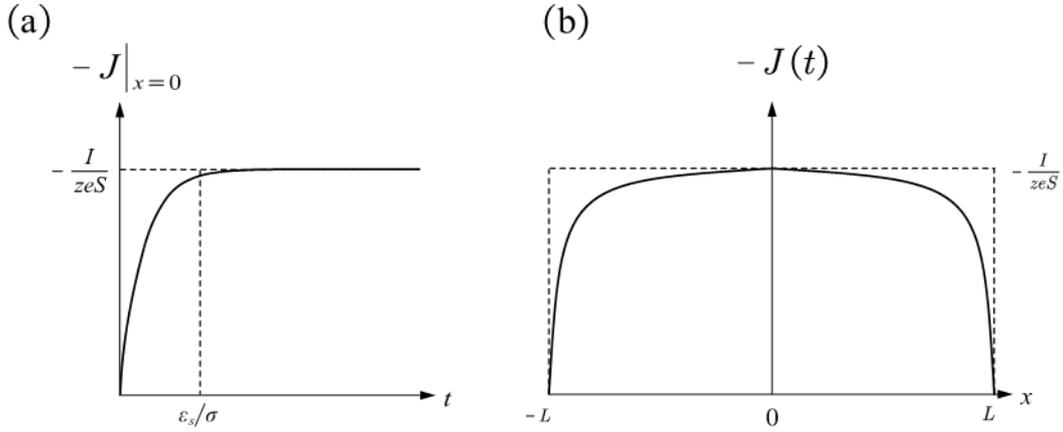

**FIG 7** (a) The function of bulk IF as a function of time, the data used for plotting the curve is represented by $1-e^{-10t}$, (b) A schematic diagram of the spatial distribution of IF at a certain moment after the relaxation time has been reached.

Since $\int_{-L}^{L} J dx$ in the device non-equilibrium capacitance (Eq. (A11)) precisely corresponds to the area of Fig 7(b), the area of this rectangle can be easily determined.

$$\left(-\int_{-L}^{L} J dx\right)_{max} = (-A)_{max} = \frac{I}{zeS}2L \quad (B10)$$

Therefore, substituting this into the non-equilibrium capacitance in EDLC (Eq. (A11)) yields the following inequality:

$$\frac{1}{C_D} + \frac{ze\int_{-L}^{L} J dx}{\varepsilon_s \, I} > 0 \quad (B11)$$

Furthermore, since the IF is opposite in direction to the current, another inequality can be obtained:

$$0 < \frac{1}{C} \leq \frac{1}{C_D} \quad (B12)$$

**APPENDIX C: IF AND THE TENDENCY OF THE SYSTEM TO RESTORE EQUILIBRIUM**

Reviewing the response process of the electrolyte to electrode charging: the electrode, through electronic current, directly increases the magnitude of the boundary electric field in the ionic system, leading to an increase in the magnitude of the electromigration flux $J_{EM}$. However, due to the electrode's obstruction of ionic transport, the resulting ion accumulation induces electrostatic repulsion among like-charged ions, which screens the external electric field, thereby impeding the further increase in electromigration flux. Additionally, the ion accumulation induces an ion concentration gradient, which generates non-uniform thermal motion and consequently gives rise to a diffusion flux $J_{diff}$. Together, the electromigration flux and the diffusion flux constitute the IF:

$$J = J_{EM} + J_{diff} \qquad (C1)$$

Since the diffusion flux is opposite in direction to the electromigration flux, it also impedes the further increase of the IF. Therefore, it can be concluded that the IF induced by charging hinders its own increase through ion accumulation (Fig 8), leading to a tendency for ion exchange within the electrolyte to restore equilibrium.

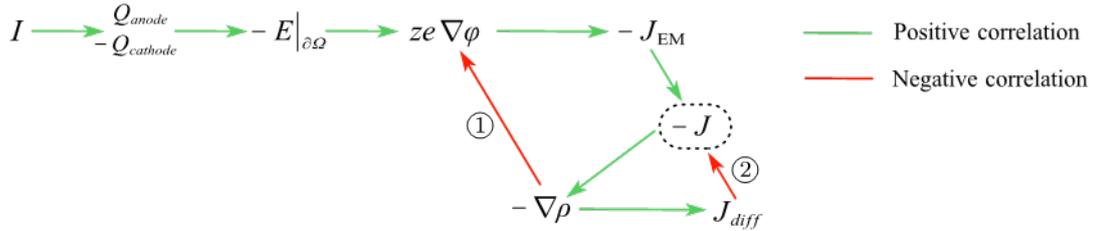

**FIG 8** Response of the electrolyte system to the charging current. (Note: the negative sign indicates the opposite direction to the current)